\documentclass[a4wide]{article}

\usepackage{graphicx}

\newcommand{\nit}{\noindent}

\newcommand{\dsp}{\displaystyle}
\newcommand{\vs}[1]{\vspace{#1 ex}}
\newcommand{\hs}[1]{\hspace{#1 em}}
\newcommand{\bfr}{\begin{flushright}}
\newcommand{\efr}{\end{flushright}}
\newcommand{\bc}{\begin{center}}
\newcommand{\ec}{\end{center}}
\newcommand{\ben}{\begin{enumerate}}
\newcommand{\een}{\end{enumerate}}

\newcommand{\be}{\begin{equation}}
\newcommand{\ee}{\end{equation}}
\newcommand{\ba}{\begin{array}}
\newcommand{\ea}{\end{array}}
\newcommand{\ct}{\cite}
\newcommand{\bit}{\bibitem}
\newcommand{\dd}[2]{\frac{\partial{#1}}{\partial{#2}}}
\newcommand{\ag}{\alpha}

\newcommand{\gam}{\gamma}
\newcommand{\del}{\delta}
\newcommand{\eps}{\epsilon}

\newcommand{\thg}{\theta}
\newcommand{\kg}{\kappa}
\newcommand{\lb}{\lambda}
\newcommand{\sg}{\sigma}

\newcommand{\fg}{\phi}
\newcommand{\vf}{\varphi}
\newcommand{\og}{\omega}
\newcommand{\Gam}{\Gamma}

\newcommand{\Fg}{\Phi}
\newcommand{\Sg}{\Sigma}

\newcommand{\cD}{{\cal D}}

\newcommand{\cH}{{\cal H}}

\newcommand{\cN}{{\cal N}}

\newcommand{\cP}{{\cal P}}
\newcommand{\cQ}{{\cal Q}}

\newcommand{\lh}{\left(}
\newcommand{\rh}{\right)}

\newcommand{\nb}{\nabla}

\newcommand{\ctg}{\mbox{\,cotan\,}}

\newcommand{\der}{\partial}

\begin{document}
\vs{3}

\nit
{\bf \large D = 1 supergravity as a constrained system}\\
\vs{2}

\hs{2} {\bf Jan W. van Holten}
\vs{1}

\hs{2} {\small Nikhef, Science Park 105, 1098XG Amsterdam, NL} 
\vs{.1}

\hs{2} {\small Lorentz Institute, Leiden University, Niels Bohrweg 2, 2333CA Leiden, NL}
\vs{.1}

\hs{2} {\small E-mail: v.holten@nikhef.nl}
\vs{1}

\nit
{\small {\bf Abstract} \\
I review the classical and quantum dynamics of systems with local world-line supersymmetry.
The hamiltonian formulation, in particular the covariant hamiltonian approach, is emphasized.
Anomalous behaviour of local quantum supersymmetry is investigated and illustrated by  
supersymmetric dynamics on the sphere $S^2$.}

\section{Introduction \label {s1}}

Standard dynamical systems are described in terms of a set of $r$ real or complex generalized 
co-ordinates $\left\{ a_i(\tau) \right\}$, $i = 1,...,r$, defining the manifold of possible configurations 
of the system. The evolution of a classical dynamical system then traces out a continuous curve 
in this configuration space, and solving the dynamics amounts to finding a prescription to construct 
this curve for given initial data. Borrowing a term from the theory of relativity, this curve is sometimes 
referred to as the {\em world line} of the system. The evolution of a quantum dynamical system 
amounts to finding a probability amplitude for the system to develop from the initial configuration 
into some specific final configuration; a heuristic way to construct such amplitudes is in terms of 
the path integral, computed by summing complex phases defined in terms of paths in the classical 
configuration space. 

Many systems of interest also possess discrete degrees of freedom, like spin or bit variables. 
Such discrete quantities may conveniently be represented by anti-commuting Grassmann variables
$\left\{ \psi_a(\tau) \right\}$, $a = 1,...., s$, with $\psi_a \psi_b = - \psi_b \psi_a$ \ct{berezin:1977}. 
For Grassmann variables a complete calculus has been developed \ct{berezin:1966} defining 
operations like differentiation and integration of Grassmann-valued functions, 
as a result of which they can be treated as pseudo-classical continuous degrees of freedom. The 
configuration space of dynamical systems involving discrete variables can then be mapped to a 
graded manifold with both classical and pseudo-classical co-ordinates $\left\{ a_i, \psi_a \right\}$. 

A special class of systems with grassmannian degrees of freedom is formed by systems with a 
symmetry relating the dynamics of the classical and pseudo-classical variables 
\ct{barducci:1976}-\ct{jwvh:1988}. This symmetry is known as world-line 
supersymmetry. In the following the dynamics of such supersymmetric systems is developed. 

\section{$D = 1$ superfields \label{s2}} 

One starting point for the construction of systems with world-line supersymmetry is to complement 
the continuous real time parameter $\tau$ with an grassmannian parameter $\thg$, $\thg^2 = 0$, 
to form a one-dimensional graded base space spanned by $(\tau, \thg)$. The most common 
degrees of freedom of the system can then be represented by real Grassmann-even functions 
\be
\Fg(\tau,\thg) = a(\tau) + i \thg \psi(\tau),
\label{2.1}
\ee
representing a graded pair $(a(\tau), \psi(\tau))$ of even and odd dynamical degrees of freedom.
The factor of $i$ has been introduced in (\ref{2.1}) as neither $\thg$ nor $\psi$ are affected by 
complex conjugation, but this operation does reverse the order of the variables. 

We consider graded translations of the world-line parameters defined by the graded pair of shift 
parameters $(\xi, \eps)$ and acting on the base-space co-ordinates as 
\be
\tau' = \tau + \xi + i \thg \eps, \hs{2} \thg' = \thg - \eps.
\label{2.2}
\ee
On the function $\Fg(\tau,\thg)$ these translations act as 
\be
\ba{lll}
\Fg(\tau', \thg') & = &  \Fg(\tau, \thg) + \xi \der_{\tau} a - i \eps \psi + i \thg \lh \xi \der_{\tau} \psi 
 + \eps \der_{\tau} a \rh + {\cal O}[\xi^2, \eps \xi] \\
 & & \\
 & = & \Fg(\tau, \thg) + \lh \xi P + i \eps Q \rh \Fg(\tau,\thg) +  {\cal O}[\xi^2, \eps \xi],
\ea
\label{2.3}
\ee
where the graded translation operators are defined by 
\be
P = \der_{\tau}, \hs{2} Q = i \der_{\thg} - \thg \der_{\tau}.
\label{2.4}
\ee
They have the algebraic properties 
\be
i Q^2 = P, \hs{2} P Q - QP = 0.
\label{2.5}
\ee
The operator $P$ is the usual time-translation operator, whilst the Grassmann-odd operator $Q$ 
is the supersymmetry operator. By itself it acts on the components in such a way that to first order 
in the shift parameters 
\be
\del a = - i \eps \psi, \hs{2} \del \psi = \eps \der_{\tau} a. 
\label{2.6}
\ee
These transformations define infinitesimal supersymmetry transformations on the graded pair
of world-line variables $(a, \psi)$. Now observe that the superderivative operator 
\be
D = i \der_{\thg} + \thg \der_{\tau}
\label{2.7}
\ee
has the properties 
\be
-i D^2 = P, \hs{2} DQ + Q D = 0, \hs{2} DP - PD = 0.
\label{2.8}
\ee
A brief calculations then shows that 
\be
I = - \frac{i}{2} \int d\tau \int d\thg\, D\Fg D^2 \Fg = 
  \frac{1}{2} \int d\tau \lh (\der_{\tau} a)^2 + i \psi \der_{\tau} \psi \rh
\label{2.9}
\ee
defines an action which is invariant under the supersymmetry transformations (\ref{2.6}) modulo
boundary terms.

\section{Local world-line supersymmetry \label{s3}} 

The supersymmetry transformations in the previous section were defined as the Grassmann-odd 
part of the constant graded world-line shifts (\ref{2.2}). In this section we introduce a set of graded 
{\em local} world-line transformations including a local realization of supersymmetry 
\ct{brink:1977,jwvh:1995,jwvh:2015}. Such a formalism is most conveniently developed in terms 
of graded pairs of variables, rather than using the $D = 1$ superfield formalism described in 
section \ref{s2}.   

As a first step consider local time reparametrizations $\tau \rightarrow \tau' = \tau - \xi(\tau)$.
As common in differential geometry we distinguish between world-line scalars $A(\tau)$ 
and world-line 1-forms $\cN = N(\tau) d\tau$ transforming as 
\be
A'(\tau') = A(\tau), \hs{2} N'(\tau')\, d \tau' = N(\tau)\, d\tau.
\label{3.1}
\ee
Following the conventions of general relativity for time-reparametrizations we introduce a specific 
1-form $\cN$ referred to as the lapse function. In terms of this a reparametrization-invariant 
derivative and an invariant integral are defined for scalars by 
\be
\cD A = \frac{1}{N}\, \frac{dA}{d\tau}, \hs{2} I = \int d\tau\, N(\tau) A(\tau).
\label{3.2}
\ee
To first order in $\xi(\tau)$ the scalar and lapse function transform as 
\be
\del A = \xi \der_{\tau} A, \hs{2} \del N = \xi \der_{\tau} N + N \der_{\tau} \xi = \der_{\tau} \lh \xi N \rh.
\label{3.3}
\ee
We now also introduce time-dependent supersymmetry transformations in terms of a 
Grassmann-odd function $\eps(\tau)$, and generalize the previous discussion to 
consider combined time- and super-reparametrizations. First we define a graded pair 
of variables $G = (N, \chi)$ involving the lapse function and a Grassmann-odd scalar\footnote{ 
Equivalently one can introduce a graded pair of 1-forms $(N, \og)$ where $\og = N \chi$; in
applications the use of $\chi$ is more convenient.} transforming as 
\be
\del N = \der_{\tau} \lh \xi N \rh - 2 i \eps \chi N, \hs{2} \del \chi = \xi \der_{\tau} \chi + \cD \eps.
\label{3.4}
\ee
These transformations obey the commutation rules 
\be
\left[ \del(\xi_2, \eps_2), \del(\xi_1, \eps_1) \right] = \del(\xi_3, \eps_3), 
\label{3.5}
\ee
with 
\be
\xi_3 = \xi_1 \der_{\tau} \xi_2 - \xi_2 \der_{\tau} \xi_1 - \frac{2i \eps_1 \eps_2}{N}, \hs{2}
\eps_3 = \xi_1 \der_{\tau} \eps_2 - \xi_2 \der_{\tau} \eps_1 + 2i \eps_1 \eps_2 \chi.
\label{3.6}
\ee
As the commutator algebra of the transformations closes they form a well-defined infinitesimal 
graded transformation group. Two more realizations of this infinitesimal group will be introduced
here. The first is one is in terms of a graded pair $\Sg = (a, \psi)$, where $a(\tau)$ is 
Grassmann-even and $\psi(\tau)$ is Grassmann-odd. On this pair the superparametrizations 
are defined by
\be
\del a = \xi \der_{\tau} a - i \eps \psi, \hs{2} 
\del \psi = \xi \der_{\tau} \psi + \eps \lh \cD a + i \chi \psi \rh.
\label{3.7}
\ee
The second one is an inversely graded pair $\Fg = (\eta, f)$ where $\eta(\tau)$ is odd and 
$f(\tau)$ is even, with transformations defined by 
\be
\del \eta = \xi \der_{\tau} \eta + \eps f, \hs{2} \del f = \xi \der_{\tau} f - i \eps \lh \cD \eta - f \chi \rh.
\label{3.8}
\ee
In both cases the transformations satisfy the commutation rules (\ref{3.5}), (\ref{3.6}).
Inversely graded pairs $\Fg$ are useful to construct super-invariant integrals, as 
\be
\del \left[ d\tau N \lh f - i \chi \eta \rh \right] = d \left[ - i \eps \eta + \xi N \lh f - i \chi \eta \rh \right], 
\label{3.9}
\ee
and therefore 
\be
I = \int d\tau\, N \lh f - i \chi \eta \rh 
\label{3.10}
\ee
is invariant modulo boundary terms.

It is possible to compose graded pairs by various simple rules. Scalar multiplication with a 
number $\lb$ is obvious:
\be
\lb \Sg = \lh \lb a, \lb \psi \rh, \hs{2} \lb \Fg = \lh \lb \eta, \lb f \rh,
\label{3.11}
\ee 
and addition and linear combination of graded pairs of same type is straightforward: 
\be
\ba{l} 
\lb \Sg_1 + \mu \Sg_2 = \lh \lb\, a_1 + \mu\, a_2,  \lb\, \psi_1 + \mu\, \psi_2 \rh, \\
 \\
\lb \Fg_1 + \mu \Fg_2 = \lh \lb\, \eta_1 + \mu\, \eta_2, \lb f_1 + \mu f_2 \rh.
\ea
\label{3.12}
\ee
Multiplication of (inversely) graded pairs labeled by $i,j, (a,b)$ is defined by the 
following rules: 
\be
\ba{l}
\Sg_i \times \Sg_j = \Sg_{ij} = \lh a_i a_j, a_i \psi_j + a_j \psi_i \rh, \\
 \\
- i \Fg_a \times \Fg_b = \Sg_{ab} = \lh - i \eta_a \eta_b, f_a \eta_b + f_b \eta_a \rh, \\
 \\
\Sg_i \times \Fg_a = \Fg_{ia} = \lh a_i \eta_a, a_i f_a - i \psi_i \eta_a \rh.
\ea
\label{3.13}
\ee
The above rules allow the construction of arbitrary polynomial functions of graded pairs, e.g.
\be
F(\Sg_i) = \lh F(a_i), \psi_j \der_j\, F(a_i) \rh.
\label{3.13a}
\ee
Finally one can introduce the superderivative $D$ acting on pairs as
\[
D: \hs{0.5} \Sg\, \stackrel{D}{\longrightarrow}\, \Fg\, \stackrel{D}{\longrightarrow}\, \Sg'\, 
  \stackrel{D}{\longrightarrow}\, \Fg'\, \stackrel{D}{\longrightarrow}\, ...
\]
by the rules 
\be
\Fg = D \Sg = \lh \psi, \cD a + i \chi \psi \rh, \hs{2} D \Fg = \lh f, \cD \eta - \chi f \rh. 
\label{3.14}
\ee
It then follows that 
\be
\Sg' = D^2 \Sg = \lh \cD a + i \chi \psi, \cD - \chi \cD a \rh, \hs{2} 
\Fg' = D^2 \Fg = \lh \cD \eta - \chi f, \cD f + i \chi \cD \eta \rh.
\label{3.15}
\ee
These components actually define {\em supercovariant derivatives}: 
\be
\ba{l} 
\nb a = \cD a + i \chi \psi, \hs{2} \nb \psi = \cD \psi - \chi \nb a, \hs{2} 
\nb^2 a - \cD \nb a + i \chi \nb \psi, \hs{2} ..., \\
 \\
\nb \eta = \cD \eta - \chi f, \hs{2} \nb f = \cD f + i \chi \nb \eta, \hs{2} ...,
\ea
\label{3.16}
\ee
with the property that $(\nb a, \nb \psi)$ and $(\nb \eta, \nb f)$ are (inversely) graded pairs 
if this holds for $(a, \psi)$ and $(\eta, f)$, respectively. Clearly there is a rule $D^2 = \nb$, 
keeping in mind that  $\nb$ is defined on individual components.

\section{Dynamics \label{s4}}

The simplest procedure to develop the dynamics of supersymmetric systems is to construct 
invariant actions. This can be done by generalization and extension of the action (\ref{2.9}) 
to representations of local supersymmetry, making use of the construction (\ref{3.10}). In this 
context we take a set of pairs $\Sg^i$, $i = 1,...,r$, interpreted as the co-ordinates of some 
graded manifold which will become the configuration space of our dynamical system. The 
first step is to form the inversely graded pairs
\be
D \Sg^i \times D^2 \Sg^j = \lh \psi^i \nb a^j, \nb a^i \nb a^j + i \psi^i \nb \psi^j \rh.
\label{4.1}
\ee
The second step is to complete this expression by contraction with a function $G_{ij}(\Sg)$ 
acting as a metric on the graded manifold: 
\be
\ba{lll}
\Fg & = & G_{ij}(\Sg) \times D \Sg^i \times D^2 \Sg^j \\
 & & \\
 & = & \dsp{ \lh G_{ij} \psi^i \nb a^j, G_{ij} \nb a^i \nb a^j  
  + i G_{ij}\, \psi^i \hs{-.1} \lh \nb \psi^j + \nb a^k \Gam_{kl}^{\;\;\,j} \psi^l \rh \rh, }
\ea
\label{4.2}
\ee
where $\Gam_{kl}^{\;\;\,j}(a)$ is the connection constructed from the metric $G_{ij}(a)$.
If we now substitute the components of $\Fg$ into the expression (\ref{3.10}) and normalize 
we get a supersymmetric action 
\be
\ba{lll}
I & = & \dsp{ \frac{1}{2}\, \int d\tau\, N \left[ G_{ij} \nb a^i \nb a^j + i G_{ij} \psi^i \lh \nb \psi^j + 
 \nb a^k \Gam_{kl}^{\;\;\,j} \psi^l \rh - i \chi\, G_{ij} \psi^i \nb a^j \right] }\\
 & & \\
 & = & \dsp{  \int d\tau\, N \left[ \frac{1}{2}\, G_{ij} \cD a^i \cD a^j + \frac{i}{2}\, G_{ij} \psi^i \lh \cD \psi^j 
  + \cD a^k \Gam_{kl}^{\;\;\,j} \psi^l \rh + i \chi\, G_{ij} \psi^i \cD a^j \right], }
\ea
\label{4.3}
\ee
transforming under local supersymmetry into
\be
\del I = \int d \lh - \frac{i\eps}{2}\, G_{ij} \hs{.1} \psi^i \cD a^j \rh.
\label{4.4}
\ee
By varying the action (\ref{4.3}) with respect to the dynamical degrees of freedom $(a^i, \psi^i)$, 
keeping the boundary values fixed, one derives the equations of motion. They can be written in 
manifestly supersymmetric form as 
\be
\ba{l}
\dsp{ \nb^2 a^i + \Gam_{jk}^{\;\;\,i}\, \nb a^j \nb a^k - \frac{i}{2}\, \psi^k \psi^l R_{klj}^{\;\;\;\;\,i} \hs{.1} \nb a^j = 0, }\\
 \\
\dsp{ \nb \psi^i + \nb a^k\, \Gam_{kj}^{\;\;\;i} \hs{.1} \psi^j = 0, }
\ea
\label{4.5}
\ee
where $R_{klj}^{\;\;\;\;\,i}(a)$ is the Riemann tensor of the configuration manifold with metric $G_{ij}(a)$. 
Furthermore, varying the action with respect to the non-dynamical variables $(N, \chi)$ one finds two 
first-class constraints related to the local time- and super-reparametrization invariance: 
\be
\cQ = G_{ij} \cD a^i \psi^j = 0, \hs{2} \cH = \frac{1}{2}\, G_{ij} \cD a^i \cD a^j = 0.
\label{4.6}
\ee
They are referred to as the supercharge and hamiltonian constraint, respectively 
\ct{teitelboim:1977, death:1988, vargas:2010}. Observe that for positive definite metrics $G_{ij}$ the 
hamiltonian constraint is extremely restrictive, essentially freezing all degrees of freedom. To get 
non-trivial dynamics an indefinite metric is strongly favored. 

\section{Hamiltonian fomulation \label{s5}}

The aim of the hamiltonian formalism is to replace a set of $r$ second-order differential equations
by $2r$ first-order differential equations. In that spirit we develop a procedure to recast the dynamics 
of the supersymmetric systems introduced in the previous section entirely in terms of first-order 
differential equations \ct{jackiw:1993}. We do this by a Legendre transform of the action for the 
Grassmann-even variables $a^i$ only, as the odd variables $\psi^i$ already obey first-order 
equations of motion \ct{jwvh:2015,jwvh:2017}. To this effect we define the Grassmann-even 
momenta $p_i$ by 
\be
p_i = \frac{\del I}{\del \dot{a}^i} = G_{ij} \lh \cD a^j + i \chi \psi^j \rh + \frac{i}{2}\, G_{ij,k} \psi^j \psi^k,
\label{5.1}
\ee
where as usual the overdot denotes a derivative with respect to time $\tau$ and we use the comma 
notation for a partial derivative w.r.t.\ any of the $a^i$. Replacing the dynamical velocities $\cD a^i$ 
in the action by the momenta it takes the form 
\be
I_c = \int d\tau\, N \lh p_i \cD a^i + \frac{i}{2}\, G_{ij} \psi^i \cD \psi^j - \cH_c \rh,
\label{5.2}
\ee
where the hamiltonian $\cH_c$ now is a function of the generalized co-ordinates and momenta:
\be
\cH_c = \frac{1}{2}\, G^{ij} \lh p_i - \frac{i}{2}\, G_{ik,l} \psi^k \psi^l + i G_{ik}\chi \psi^k \rh 
 \lh p_j - \frac{i}{2}\, G_{jm,n} \psi^m \psi^n + i G_{jn}\chi \psi^n \rh.
\label{5.3}
\ee
It is easy to verify that a similar substitution of momenta in the supercharge results in 
\be
\cQ_c = p_i \psi^i.
\label{5.4}
\ee
The action (\ref{5.2}) with the hamiltonian (\ref{5.3}) is guaranteed to produce equations of motion 
for the dynamical degrees of freedom which are first-order differential equations in time: 
\be
 \lh \ba{ccc}
  0 & - \del_i^j & \frac{i}{2} G_{kj,i} \psi^k \\
  & & \\ & & \\
  \del_j^i & 0 & 0 \\
  & & \\ & & \\
  \frac{i}{2} G_{ik,j} \psi^k & 0 & i G_{ij} \ea \rh 
  \left[ \ba{c}  \cD a^j \\ \\ \\ \cD p_j \\ \\ \\ \cD \psi^j \ea \right] = \left[ \ba{c}
  \dsp{ \dd{\cH}{a^i} }\\ \\ \dsp{ \dd{\cH}{p_i} }\\  \\ \dsp{ \dd{\cH}{\psi^i} }\ea \right].
\label{5.5}
\ee
These equations can be inverted to give 
\be
 \left[ \ba{c}  \cD a^i \\ \\ \\ \cD p_i \\ \\ \\ \cD \psi^i \ea \right] = 
 \lh \ba{ccc}
  0 & \del^i_j & 0 \\
  & & \\ & & \\
 - \del_i^j & - \frac{i}{4} G^{mn} G_{mk,i} G_{nl,j} \psi^k \psi^l & \frac{1}{2} G_{kl,i} G^{lj} \psi^k \\
  & & \\ & & \\
  0 &  - \frac{1}{2} G^{il} G_{lk,j} \psi^k & -i G^{ij}  \ea \rh 
\left[ \ba{c} 
  \dsp{ \dd{\cH}{a^i} }\\ \\ \dsp{ \dd{\cH}{p_i} }\\  \\ \dsp{ \dd{\cH}{\psi^i} } \ea \right]. 
\label{5.6}
\ee
The canonical hamiltonian formulation of the dynamics now amounts to the following: there exists
a graded continuum of (gauge-equivalent) phase spaces labeled by the graded pair $(N, \chi)$ of 
which the dynamical variables $(a^i, p_i, \psi^i)$ are the co-ordinates. For any specific choice of
$(N, \chi)$ the dynamics on that representative phase space is defined by a bracket 
\be
\cD F = \left\{ F, \cH_c \right\},
\label{5.7}
\ee
generating the evolution equations of a phase-space function $F(a^i, p_i, \psi^i)$ for fixed 
$(N, \chi)$ by application of the canonical brackets
\be
\ba{ll}
\dsp{ \left\{ a^i, p_j \right\}  = - \left\{ p_j, a^i \right\} = \del_j^i, }& \dsp{
\left\{ p_i, p_j \right\} = - \left\{ p_j, p_i \right\} = - \frac{i}{4}\, G^{mn} G_{mk,i} G_{nl,j} \psi^k \psi^l, }\\ 
 & \\
\dsp{ \left\{ \psi^i, \psi^j \right\} = \left\{ \psi^j, \psi^i \right\} = - i G^{ij}, }& \dsp{ 
\left\{ p_i, \psi^j \right\} = - \left\{ \psi^j, p_i \right\} = \frac{1}{2}\, G_{kl,i} G^{lj} \psi^k. }
\ea
\label{5.8}
\ee
The dynamical equations (\ref{5.7}) thus represent the appropriate generalization of the Hamilton 
equations to locally supersymmetric systems. An obvious choice of representative phase space 
it the one labeled by $(N, \chi) = (1,0)$; however these values are to be substituted only after 
imposing the first-class constraints (\ref{4.6}) in the form 
\be
\cQ_c = 0, \hs{2} \cH_c = 0.
\label{5.9}
\ee
We close this section by observing that 
\be
\left\{ \cQ_c, \cQ_c \right\} = 2 i \lh \cH_c - i \chi \cQ_c \rh, \hs{2} 
\left\{ \cQ_c, \cH_c \right\} = 2 \chi \cH_c.
\label{5.10}
\ee
in agreement with the first-class nature of the constraints. 

\section{Hamilton-Jacobi equation \label{s6}}

Consider the solutions of the equations of motion tracing out curves in phase space between 
a fixed initial point at time $\tau_1$ and various end points $(a^i, p_i, \psi^i)$ at time $\tau_2$. 
Then one can define a function of the end point $S(a^i,\psi^i)$ by integrating the action $I_c$ 
over the corresponding curve: 
\be
S = \int_{\tau_1}^{\tau_2} d\tau\, N \lh p_i \cD a^i + \frac{i}{2}\, G_{ij} \psi^i \cD \psi^j - \cH_c \rh.
\label{6.1}
\ee
To see that this is a function only of $(a^i, \psi^i)$ it suffices to consider variations of the end 
point with different corresponding paths in phase space. As along each path the equations of 
motion (\ref{5.5}) are satisfied the corresponding variation of $S$ comes from the variation 
of the boundary point and is given (for fixed initial point) by 
\be
\del S =  \left[ \del a^i p_i - \frac{i}{2}\, \del \psi^i G_{ij} \psi^j \right]_{\tau_2}.
\label{6.2}
\ee
There is no contribution from variations $\del p_i$, as the time-derivatives of the momenta do 
not appear in the action or in the integral (\ref{6.1}). As the variations $\del a^i$ and $\del \psi^i$
at the end point are independent, it also follows that the dependence of $S$ on the end points 
results in 
\be
\dd{S}{a^i} = p_i, \hs{2} \dd{S}{\psi^i} = - \frac{i}{2}\, G_{ij} \psi^j.
\label{6.3}
\ee
In view of this the first-class constraint for the supercharge, which is satisfied for all true 
solutions, implies that 
\be
G^{ij} \dd{S}{a^i} \dd{S}{\psi^j} = 0.
\label{6.4}
\ee
This is the supersymmetric variant of the Hamilton-Jacobi equation \ct{jwvh:2015}. 

\section{Covariant hamiltonian formalism \label{s7}}

The equations of motion (\ref{4.5}) have a purely geometric form, describing the world line of 
a spinning particle in $r$ dimensions on which the spin variables $\psi^i$ move by parallel 
transport. In the canonical hamiltonian formulation, obtained in section \ref{s5} by Legendre 
transform of the same action (\ref{4.3}), the manifest geometric structure is lost. In particular 
the canonical brackets (\ref{5.8}) do not have a direct geometric representation. Manifestly
covariant formulations of the dynamics do however exist \ct{jwvh:2006}. In this section we 
develop actually two such formulations. 

The first one is obtained by replacing the canonical momenta by covariant momenta: 
\be
p_i\; \rightarrow\; \cP_i = p_i - \frac{i}{2}\, G_{ij,k} \psi^j \psi^k.
\label{7.1}
\ee
They are covariant as after reparametrizing phase space in terms of $(a^i, \cP_i, \psi^i)$ 
the brackets (\ref{5.8}) are replaced by the manifestly covariant expressions
\be
\ba{ll}
\left\{ a^i, \cP_j \right\} = - \left\{ \cP_j, a^i \right\} = \del_j^i, & \dsp{
\left\{ \cP_i, \cP_j \right\} = - \left\{ \cP_j, \cP_i \right\} = - \frac{i}{2}\, \psi^k \psi^l R_{klij}, }\\
 & \\ 
\left\{ \psi^i, \psi^j \right\} = \left\{ \psi^j, \psi^i \right\} = - i G^{ij}, & 
\left\{ \cP_i, \psi^j \right\} = - \left\{ \psi^j, \cP_i \right\}  = \Gam_{ik}^{\;\;\,j} \psi^k,
\ea
\label{7.2}
\ee
where the structure functions of the various brackets are defined by the metric, the connection 
and the Riemann curvature of the manifold \ct{ambrosi:2015}.

In addition to a transformation of the phase-space co-ordinates and brackets, to reobtain 
the covariant equations of motion (\ref{4.5}) we also recast the supercharge and hamiltonian
in covariant form:
\be
\cQ_{cov} = \cP_i \psi^i, \hs{2} \cH_{cov} = \frac{1}{2}\, G^{ij} \cP_i \cP_j.
\label{7.3}
\ee
These are obtained from $\cQ_c$ and $\cH_c$ by the substitution (\ref{7.1}) and subsequently 
taking
\be
\cQ_{cov} = \cQ_c , \hs{2} \cH_{cov} = \cH_c + i \chi \cQ_c.
\label{7.4}
\ee
Note that such a recombination of $\cH_c$ and $\cQ_c$ does not alter the first-class constraints,
as they are equivalent with
\be
\cQ_{cov} = 0, \hs{2} \cH_{cov} = 0.
\label{7.5}
\ee
However, the canonical equations of motion (\ref{5.7}) are now replaced by the covariant 
Hamilton equations 
\be
\nb F = \cD F + i \chi \left\{ F, \cQ_{cov} \right\} = \left\{ F, \cH_{cov} \right\},
\label{7.6}
\ee
for any function $F(a, \cP, \psi)$ on the phase space. To evaluate these expressions we collect 
here the phase-space supersymmetry transformations generated by the supercharge 
\be
\left\{ a^i, \cQ_{cov} \right\} = \psi^i, \hs{2} \left\{ \cP_i, \cQ_{cov} \right\} = \Gam_{ij}^{\;\;\,k} \cP_k \psi^j, 
 \hs{2} \left\{ \psi^i, \cQ_{cov} \right\} = - i G^{ij} \cP_j.
\label{7.7}
\ee
It then follows in particular that 
\be
\cP_i = G_{ij} \nb a^j. 
\label{7.8a}
\ee
Using this expression in the equations of motion for $\cP_i$ and $\psi^i$ returns the covariant 
equations (\ref{4.5}). As concerns the algebra of constraints, with the help of the relations (\ref{7.7}) 
it is straighforward to establish that 
\be
\left\{ \cQ_{cov}, \cQ_{cov} \right\} = - 2i \cH_{cov}, \hs{2} 
\left\{ \cQ_{cov}, \cH_{cov} \right\} = 0.
\label{7.8}
\ee
Clearly these relations are in full agreement with the constraints (\ref{7.5}), confirming again their 
first-class character. 

Another covariant formulation, fully equivalent at the level of classical dynamics, can be constructed 
in terms of local tangent frames. These are spanned at any point of the manifold by an orthonormal 
set of $r$ vectors $e^i_a(a)$:
\be
G_{ij}\, e^i_a e^j_b = \eta_{ab}, \hs{2} a,b = (1,...,r),
\label{7.9}
\ee
where $\eta_{ab}$ is the tangent-space euclidean or pseudo-euclidean metric with all eigenvalues 
$\pm 1$; in locally (pseudo-)cartesian co-ordinates $\eta =$ diag$(\pm 1, ..., \pm 1)$. Being 
non-singular there exist dual 1-forms $e^a = e_i^a(a)\, da^i$ such that 
\be
e^i_a e_i^b = \del_a^b,
\label{7.10}
\ee
from which it follows that 
\be
e_i^a = G_{ij}\, e_b^j\, \eta^{ba}, \hs{2} \eta_{ab}\, e_i^a e_j^b = G_{ij}.
\label{7.11}
\ee
Parallel transport of the frames now involves local $r$-dimensional (pseudo-)rotations in the tangent 
space, which are kept track of by the spin connection $\og^a_{\;\,b} = \og_{i\;\,b}^{\;a}(a)\, da^i$:
\be
e_{j\; ;i}^a = \der_i e_j^a - \Gam_{ij}^{\;\;\,k} e_k^a = \og_{i\;\,b}^{\;a}\hs{.1} e_j^b.
\label{7.12}
\ee
Representing a (pseudo-)rotation the spin connection has the anti-symmetry property
\[
\og_{ab} = \eta_{ac}\, \og^c_{\;\,b} = - \og_{ba}.
\] 
From the definition (\ref{7.12}) and the Ricci identity it is straightforward to establish that the 
covariant field strength of the spin connection is directly related to the Riemann tensor by
\be
R_{ijab} = \der_i \og_{jab} - \der_j \og_{iab} - \left[ \og_i, \og_j \right]_{ab} 
 = R_{ijkl}\, e^k_a e^l_{\hs{.1}b} = R_{ab ij}.
\label{7.13}
\ee
In the context of supersymmetric dynamical systems we use the tangent frames to redefine 
the Grassmann-odd co-ordinates as taking values in the local tangent frame: 
\be
\psi^i\; \rightarrow\; \fg^a = e_i^a \psi^i.
\label{7.14}
\ee
The brackets (\ref{7.2}) then are replaced by 
\be
\ba{ll}
\left\{ a^i, \cP_j \right\} = - \left\{ \cP_j, a^i \right\} = \del_j^i, & \dsp{
\left\{ \cP_i, \cP_j \right\} = - \left\{ \cP_j, \cP_i \right\} = - \frac{i}{2}\, \fg^a \fg^b R_{abij}, }\\
 & \\ 
\left\{ \fg^a, \fg^b \right\} = \left\{ \fg^b, \fg^a \right\} = - i \eta^{ab}, & 
\left\{ \cP_i, \fg^a \right\} = - \left\{ \fg^a, \cP_i \right\}  = - \og_{i\;\,b}^{\;a} \fg^b.
\ea
\label{7.15}
\ee
After the redefinition the supercharge becomes
\be
\cQ_{cov} = \cP_i e_a^i \fg^a.
\label{7.16}
\ee
The hamiltonian (\ref{7.3}) and the algbera of constraints (\ref{7.8}) remain unchanged.
The equations of motion now read 
\be 
\ba{l}
\cP_i = G_{ij} \nb a^j, \hs{2} \dsp{
 \nb^2 a^i + \Gam_{jk}^{\;\;\,i} \nb a^j \nb a^k = \frac{i}{2}\, \fg^a \fg^b R^{\;\;\;\;\;\,i}_{abj} \nb a^j, }\\
 \\
\dsp{ \nb \fg^a + \nb a^i\, \og_{i\;\,b}^{\;a}\, \fg^b = 0. }
\ea
\label{7.17}
\ee

\section{Quantum theory \label{s8}}

The local tangent-frame formulation of the pseudo-classical dynamics of locally supersymmetric 
systems is most convenient for the purpose of canonical quantization \ct{jwvh:2015}. In this 
procedure the phase-space variables are replaced by self-adjoint operators in Hilbert space, and 
the correspondence principle is used to determine the commutation or anti-commutation relations 
between these operators from the classical brackets. For the systems at hand the first step is 
to introduce operators corresponding to the physical degrees of freedom 
\be
a^i \rightarrow \xi^i, \hs{2} \cP_i \rightarrow \pi_i, \hs{2} \fg^a \rightarrow \frac{1}{\sqrt{2}}\, \gam^a.
\label{8.1}
\ee
Following the rule 
\[
i \mbox{(phase space brackets)} \hs{.5} \rightarrow \hs{.5} \left[ \mbox{(quantum operators)} \right\} ,
\]
with square brackets $[~,~]$ denoting commutators and accolades $\{ ~, ~\}$ now anti-commutators, 
the fundamental operator commutation relations are postulated to be
\be
\ba{ll}
\left[ \xi^i, \pi_j \right] = i \del_j^i, & \left\{ \gam^a, \gam^b \right\} = 2 \eta^{ab}, \\
 & \\
\left[ \gam^a, \pi_i \right] = i \og_{i\;\,b}^{\;a} \gam^b, & \dsp{ 
\left[ \pi_i, \pi_j \right] = \frac{1}{2}\, \sg^{ab} R_{abij}, }
\ea
\label{8.2}
\ee
with $\gam^a$ acting as the generators of an $r$-dimensional Clifford algebra and  
\be
\sg^{ab} = \frac{1}{4} \left[ \gam^a, \gam^b \right] 
\label{8.3}
\ee
as the generators of the associated (pseudo) rotation group $SO(r-s, s)$. Here $s$ 
denotes the number of non-compact (time-like) dimensions in tangent space. 
The consistency of the commutation relations (\ref{8.2}) may be checked from the 
graded Jacobi identities, using the Clifford and Bianchi identities
\be
\left[ \sg^{bc}, \gam^a \right] = \eta^{ac} \gam^b - \eta^{ab} \gam^c, \hs{2} 
R_{ab[ij;k]} = 0,
\label{8.4}
\ee
which guarantee that for all $(A,B,C) \in \{ \xi^i, \pi_i, \gam^a\}$
\be
(-1)^{CA} \left[ \left[ A, B \right\}, C \right\} + (-1)^{AB} \left[ \left[ B, C \right\}, A \right\} 
   + (-1)^{BC} \left[ \left[ C, A \right\}, B \right\}  = 0.
\label{8.5}
\ee 
It now remains to find a realization of these operators in some representation of Hilbert space. 
The commutation relation of the $\xi^i$ and $\pi_i$ implies that the $\pi_i$ act as a set  of 
non-commutative derivatives on functions of the $\xi^i$. Next considering representations of 
the Clifford algebra spanned by the operators $\gam^a$, we observe that there is an irreducible 
representation of their anti-commutation relations in terms of Dirac matrices of dimension 
$2^{[r/2]} \times 2^{[r/2]}$, which suggests to take the elements of Hilbert space to be 
$2^{[r/2]}$-component spinors $\Psi$. However, to guarantee the self-adjointness of the 
operators some additional steps are necessary. The precise steps depend on the number $s$
of non-compact directions in tangent space. We restrict our discussion to the cases $s = 0$ 
and $s = 1$ corresponding to euclidean and minkowskian tangent-space geometries, 
respectively. 

For euclidean tangent spaces with positive-definite metric $\eta_{ab} = \del_{ab} =$ 
diag$(+1, .., +1)$ the irreducible representations of the Dirac matrices are hermitean; 
therefore one can define an invariant scalar product on the Hilbert space of spinors by 
\be
\lh \Fg, \Psi \rh = \int d^r \xi\, e\, \Fg^{\dagger} \Psi,
\label{8.6}
\ee
where $e = \det e_i^a = \sqrt{G}$ is included to make the integration measure invariant 
under co-ordinate transformations. Owing to the hermitean property if the Dirac matrices 
they represent self-adjoint operators:
\be
\lh \Fg, \gam^a \Psi \rh = \lh \gam^a \Fg, \Psi \rh.
\label{8.7}
\ee
For minkowskian tangent spaces with indefinite metric $\eta_{ab} =$ diag$(-1,+1, ..., +1)$
only  the compact (space-like) components of $\gam^a$, $a = 1, ..., r-1$, are represented 
by hermitean matrices, whilst $\gam^0$ is anti-hermitean. However, it then follows that 
one can construct a complete set of hermitean matrices $\gam^0 \gam^a$ for all
$a = 0, 1, ..., r-1$:
\be
\lh \gam^0 \gam^a \rh^{\dagger} = \gam^0 \gam^a
\label{8.8}
\ee
Following Dirac we therefore define a modified invariant scalar product 
\be
\lh \Fg, \Psi \rh = \int d^r \xi\, e \, \bar{\Fg} \Psi, \hs{2} \bar{\Fg} = \Fg^{\dagger} \gam^0,
\label{8.9}
\ee
where now $e = \sqrt{ -G}$. In view of (\ref{8.8}) it is then again guaranteed that the operators 
$\gam^a$ are self-adjoint as in (\ref{8.7}) with respect to this modified scalar product. 

Having established the conditions for the operators $\gam^a$ to be self-adjoint we can also 
provide a self-adjoint representation of the momentum operators $\pi_i$ in terms of the spin
connection:
\be
\pi_i = -\frac{i}{\sqrt{e}}\, D_i \sqrt{e}. \hs{2} D_i = \der_i - \frac{1}{2}\, \og_{iab} \sg^{ab},
\label{8.10}
\ee
as in euclidean space $\sg_{ab}^{\dagger} = - \sg_{ab}$, whilst in the minkowskian case
eq.\ (\ref{8.8}) implies that 
\be
\gam^0 \sg^{ab} = - \lh\gam^0 \sg^{ab} \rh^{\dagger}.
\label{8.11}
\ee
The first equation (\ref{8.4}) then esablishes the commutation relation between $\gam^a$ 
and $\pi_i$, whilst the commutator of two momenta holds because of the Ricci identity:
\be
\left[ \pi_i, \pi_j \right] = - \frac{1}{\sqrt{e}} \left[ D_i, D_j \right] \sqrt{e} = \frac{1}{2}\, \sg^{ab} R_{abij}.
\label{8.12}
\ee
With the operator representations defined above we can construct a self-adjoint supercharge 
operator: 
\be
\cQ = \frac{1}{2} \lh \gam^a e_a^i \pi_i + \pi_i e^i_a \gam^a \rh = - i \gam^a e_a^i D_i,
\label{8.13}
\ee
where in the last step we have used the definition of the spin-connection (\ref{7.12}). 
The supersymmetry constraint then reduces to the condition 
\be
- i \gam^a e_a^i D_i \Psi = 0, 
\label{8.14}
\ee
which is the massless Dirac equation on the curved manifold with metric $G_{ij}[\xi]$. 
The corresponding hamiltonian constraint is 
\be
\cH = \frac{1}{2} \left\{ \cQ, \cQ \right\} = - \frac{1}{2}\, D_i^2 - \frac{1}{8}\, R, 
\label{8.15}
\ee
where $R$ is the Riemann scalar. Note that any operator-ordering ambiguities in the definition 
of the hamiltonian have been solved by its relation to the supercharge and the condition of
self-adjointness of the latter \ct{jwvh:1988}. As in the classical models the consistency of the 
constraints follows from (\ref{8.15}) and its consequence
\be
\left[ \cQ , \cH \right] = 0.
\label{8.16}
\ee

\section{Examples: supersymmetric $S^2$ \label{s9}}

The evolution of dynamical quantum operators $F(\xi, \pi, \gam)$ are obtained by straighforward 
application of the hamiltonian:
\be
\cD F + i \chi \left[ F, \cQ \right\} = \left[ F, \cH \right].
\label{9.1a}
\ee
Here the auxiliary gauge variables $(N, \chi)$ are to be considered external parameters fixing 
the representation of the Hilbert space. As alluded to above, it is perfectly allowed to chose a 
particular representative, the one with $N = 1$ and $\chi = 0$ being the simplest one to work 
with. This choice is employed below. It should however be remembered that their role was to 
impose the first-class constraints, in the quantum theory on the states in Hilbert space: 
\be
\cQ \Psi = 0, \hs{2} \cH \Psi = 0.
\label{9.2a}
\ee
As the second constraint is an automatic consequence of the first one, the supercharge constraint 
is the fundamental constraint, the hamiltonian one is used for mathematical simplifications. 
Thus the physical states of a quantum system with world-line supersymmetry are characterized 
by the generalized Dirac equation (\ref{8.14}); the space of solutions consists of the set
of zero modes, the kernel, of the Dirac operator.

The existence of physical states is not guaranteed; if the kernel of the Dirac operators is empty 
there is no supersymmetric system in which the dynamical constraint can be realized. This 
might be interpreted as an anomaly of the local world-line supersymmetry at the quantum 
level. In fact we have already noticed in section \ref{s4} that in the classical theory for 
positive-definite metrics $G_{ij}$ the system is frozen, and this is manifested by the absence
of normalizable zero-modes in the quantum theory.

An explicit example of such a situation is provided by the supersymmetric sphere $S^2$
\ct{camporesi:1995}.  Its tangent space is euclidean $R^2$ and in polar co-ordinates 
$\xi^i = (\thg, \vf)$ the metric and corresponding tangent frame vectors are  
\be
G_{ij} = \lh \ba{cc} 1 & 0 \\ 0 & \sin^2 \thg \ea \rh, \hs{2} 
e_i^a = \lh \ba{cc} 1 & 0 \\ 0 & \sin \thg \ea \rh.
\label{9.1}
\ee
The tangent-space components of the spin connection 1-form are 
\be
\og_i d\xi^i = \lh \ba{cc} 0 & \cos \thg \\ - \cos \thg & 0 \ea \rh d \vf,
\label{9.2}
\ee
In two dimensions the Dirac matrices are identical with the Pauli matrices: $\gam^a = \sg^a$, 
$a = (1,2)$. The Dirac operator in this representation is then found to be 
\be
- i \gam^a e_a^i D_i = \lh \ba{cc} 0 & - i D_- \\ - i D_+ & 0 \ea \rh, \hs{2} 
D_{\pm} = \der_{\thg} + \frac{1}{2}\, \ctg \thg \pm \frac{i}{\sin \thg}\, \der_{\vf}.
\label{9.3}
\ee
To find the eigenvalues of this operator solve the equations
\be
- i \gam^a e_a^i D_i \Psi = \kg \Psi, \hs{2} \Psi(\thg,\vf) = e^{im\vf} \left[ \ba{l} \psi_+(\thg) \\ \psi_-(\thg) \ea \right], 
\label{9.4}
\ee
where $m$ is an integer; the components satisfy
\be
\dsp{ - i \lh  \der_{\thg} + \frac{1}{2}\, \ctg \thg + \frac{m}{\sin \thg} \rh \psi_- = \kg \psi_+, \hs{1} 
 - i \lh  \der_{\thg} + \frac{1}{2}\, \ctg \thg - \frac{m}{\sin \thg} \rh \psi_+ = \kg \psi_-. }
\label{9.5}
\ee
The hamiltonian form is more practical as it diagonalizes the equations:
\be 
\lh \der_{\thg}^2 + \ctg \thg\, \der_{\thg} - \frac{m^2 + \frac{1}{4}}{\sin^2 \thg} \pm
 \frac{m \cos \thg}{\sin^2 \thg} - \frac{1}{4} \rh \psi_{\pm} = - \kg^2 \psi_{\pm}.
\label{9.6}
\ee
After a change of variable $z = \cos \thg$ this equation takes the form
\be
\left[ (1- z^2)\, \der^2_{z} - 2\xi \der_z - \frac{m^2 + \frac{1}{4}}{1-z^2}\, \pm \frac{m z}{1- z^2} 
 - \frac{1}{4} \right] \psi_{\pm} = - \kg^2 \psi_{\pm}.
\label{9.7}
\ee
In the case of $m \geq 0$ we first take the upper component; the solutions are of the form
\be
\psi_+ = \lh 1 - z \rh^{\frac{m}{2} - \frac{1}{4}} \lh 1 + z \rh^{\frac{m}{2} + \frac{1}{4}} J_{n}^{m-1/2, m+1/2}(z),
\label{9.8}
\ee
where $J_n^{p,q}$ is a Jacobi polynomial of degree $n$. For $p = m -1/2$ and $q = m+1/2$ 
the corresponding eigenvalues are
\be
\kg_{nm}^2 = \lh n + m + \frac{1}{2} \rh^2.
\label{9.9}
\ee
The corresponding solutions for the lower component in (\ref{9.7}) are obtained by computing
\be
\kg \psi_- = \lh \sqrt{1 - z^2}\, \der_z - \frac{1}{2}\, \frac{z}{\sqrt{1- z^2}} 
 + \frac{m}{\sqrt{1 - z^2}} \rh \psi_+,
\label{9.10}
\ee
and have the same eigenvalue spectrum. For negative $m$ the situation is reversed, the lower 
components $\psi_-$ being described by (\ref{9.8}) and the upper components by (\ref{9.10}).
As both $n$ and $m$ are integers it follows that there are no solutions of eq.\ (\ref{9.4}) with 
$\kg = 0$. 

The same eigenvalue problem in the context of a space with indefinite metric appears 
in a cosmological setting  \ct{jwvh:2015} if we consider a homogeneous space-time of 
Friedmann-Lemaitre type with scale factor $a(t) = e^{\xi^0/\sqrt{6}}$ and two spatially 
homogeneous scalar fields $\xi^m(t)$, $m = (1,2)$ taking values on the sphere $S^2$. 
The non-supersymmetric classical action for such a system is 
\be
S = \frac{1}{2}\, \int d^4 \tau\, N\, g_{ij}\, \cD \xi^i \cD \xi^j, \hs{2} 
g_{ij} = \lh \ba{cc} -1 & 0 \\ 0 & G_{mn} \ea \rh,
\label{9.11}
\ee
where $i,j = (0,1,2)$, $m,n = (1,2)$ and $G_{mn}$ is the scalar-field metric on $S^2$. 
Introducing a cosmological time $\sg$ defined in terms of the scale factor and polar 
co-ordinates such that $\xi^i = (\sg, \thg, \vf)$ the metric reads
\be
g_{ij} = \lh \ba{ccc} -1 & 0 & 0 \\ 0 & 1 & 0 \\ 0 & 0 & \sin^2 \thg \ea \rh.
\label{9.12}
\ee
The supersymmetric version of the corresponding quantum cosmology in which 
$\gam^0 = i \sg^3$, $\gam^{1,2} = \sg^{1,2}$, is defined by the Dirac operator
\be
- i \gam^a e_a^i D_i = \lh \ba{cc} \der_{\sg} & - i D_- \\ - i D_+ & - \der_{\sg} \ea \rh,
\label{9.13}
\ee
where the covariant derivatives $D_{\pm}$ on the sphere are defined as before, eq.\ (\ref{9.3}).
The eigenfunctions are 2-component spinors 
\be
\Psi(\sg, \thg, \vf) = e^{i\kg \sg + i m \vf} \left[ \ba{c} \psi_+(\thg) \\ \psi_-(\thg) \ea \right],
\label{9.14}
\ee
and now the zero-modes are the solutions of (\ref{9.7}) for all allowed values of $\kg$. 
Thus we have an infinite set of solutions of then form (\ref{9.8}), (\ref{9.10}) labeled by 
integers $(n,m)$ with the spectrum of allowed values $\kg_{nm}$ given by (\ref{9.9}). 

\section{Potentials \label{s10}}

In the previous sections we have focussed on the description of supersymmetric dynamical 
systems with purely geometrical hamiltonians (\ref{4.6}). In this section we describe how to 
extend the dynamics with interactions of potential type \ct{jwvh:1995}. From sect.\ \ref{s3} 
we recall that supersymmetric actions are constructed from elementary or composite inversely 
graded superpairs $(\eta, f)$ using expression (\ref{3.10}). A straightforward generalization of 
this construction is to take an elementary pair $\Fg = (\eta, f)$ and multiply it with some function 
$W[\Sg]$ of the scalar pairs $\Sg_i = (a_i, \psi_i)$ according to the last rule (\ref{3.13}):
\be
\Fg_W \equiv W[\Sg] \times \Fg = \lh W(a) \eta, W(a) f - i \der_i W(a)\, \psi^i \eta \rh.
\label{10.1}
\ee
Another constribution is constructed by taking 
\be
D \Fg \times \Fg = \lh f \eta, f^2 + i \eta \nb  \eta  \rh.
\label{10.2}
\ee
Combine these terms to form 
\be
\Fg_{pot} = \frac{1}{2}\, D \Fg \times \Fg - \Fg_W, 
\label{10.3}
\ee
and insert the components of $\Fg_{pot}$ into expression (\ref{3.10}) to get 
\be
I_{pot} = \int d\tau\, N \left[ \frac{i}{2}\, \eta \cD \eta + \frac{1}{2}\, f^2 - W f - i\eta \chi W 
  - i \eta \psi^i \der_i W \right]. 
\label{10.4}
\ee
Now the auxiliary variable $f$ can be eliminated by its algebraic equation of motion $f = W$,
equivalent to completing the square, to get 
\be
I_{pot} \simeq \int d\tau\, N \left[ \frac{i}{2}\, \eta \cD \eta - i \eta \chi W - i \eta \psi^i \der_i W
 - \frac{1}{2}\, W^2 \right]. 
\label{10.5}
\ee
Adding this to the action (\ref{4.3}) it supplies a scalar potential $W^2(a) /2$ plus supersymmetric completion
terms, at the price of having an extra Grassmann-odd degree of freedom. Of course an arbitrary number of
such potentials can be added in principle, each carrying its own Grassmann-odd variable along. With the 
single contribution of the potential terms (\ref{10.5}) the equations of motion are modified to
\be
\ba{l}
\dsp{ \nb^2 a^i + \Gam_{jk}^{\;\;\,i}\, \nb a^j \nb a^k - \frac{i}{2}\, \psi^k \psi^l R_{klj}^{\;\;\;\;\,i} \hs{.1} \nb a^j 
  = - G^{ij} \lh W_{,j} W + i \eta \psi^k W_{,jk} + i \eta \chi W_{,j} \rh, }\\
 \\
\dsp{ \nb \psi^i + \nb a^k\, \Gam_{kj}^{\;\;\;i} \hs{.1} \psi^j =  G^{ij} W_{,j} \eta, }\\
 \\
\dsp{ \nb \eta = \cD \eta - \chi W = \psi^i W_{,i}, }
\ea
\label{10.6}
\ee
where as before we use the comma notation to denote partial derivatives w.r.t.\ the $a^i$. In addition we 
get modified constraints from variations w.r.t.\ $N$ and $\chi$:
\be
\ba{l}
\dsp{ \cH = \frac{1}{2}\, G_{ij} \cD a^i \cD a^j + \frac{1}{2}\, W^2 + i \eta \lh \chi W + \psi^i W_{,i} \rh = 0, }\\
 \\
\dsp{ \cQ = G_{ij} \cD a^i \psi^j + \eta W = 0. }
\ea
\label{10.7}
\ee
Upon quantization the additional Grassmann-odd variables will  become operators 
$\eta \rightarrow \ag/\sqrt{2}$ extending the Clifford algebra (\ref{8.2}) with one or more 
extra generators $\ag$: 
\be
\ag^2 = 1, \hs{2}  \ag \gam^a + \gam^a \ag = 0.
\label{10.8}
\ee
The supercharge operator (\ref{8.13}) is accordingly generalized to 
\be
\cQ = - i \gam^a e_a^i D_i + \ag W.
\label{10.9}
\ee
This allows for turning the supersymmetry constraint into version of the Dirac equation 
extended by a mass term. In the example of the sphere $S^2$ the introduction of a mass 
term in (\ref{9.3}) would be equivalent to consider a single mode of the cosmological 
model (\ref{9.13}), which we have seen to have normalizable solutions. This amounts to 
a form of reduction from 2+1 to 2 dimensions, though not by compactification but by mode 
selection. Equivalently, the full set of solutions of the cosmological model corresponds to 
a full tower of Dirac equations for massive states on $S^2$, with quantized masses (\ref{9.9}).
Thus the anomalous behaviour of local world-line supersymmetry is cured by the introduction 
of a mass term accompanied by an extension of the Clifford algebra, signifying an extra 
Grassmann-odd degree of freedom in the corresponding pseudo-classical model.

\end{document}